# Spin pumping and large field-like torque at room temperature in sputtered amorphous WTe$_{2-x}$ films

Yihong Fan[1,†], Hongshi Li[1,†], Mahendra DC[3,†,‡], Thomas Peterson[3,†], Jacob Held,[2] Protyush Sahu[3], Junyang Chen[1], Delin Zhang[1], Andre Mkhoyan[2], and Jian-Ping Wang[1,2,3]*

[1]Electrical and Computer Engineering Department, University of Minnesota
[2]Chemical Engineering and Materials Science Department, University of Minnesota
[3]School of Physics and Astronomy, University of Minnesota
*Email: jpwang@umn.edu; corresponding author
†These authors contribute equally to the work

**Abstract:**

We studied the spin-to-charge and charge-to-spin conversion at room temperature in sputtered WTe$_{2-x}$ (x=0.8)(t)/Co$_{20}$Fe$_{60}$B$_{20}$(6 nm) heterostructures. Spin pumping measurements were used to characterize the spin-to-charge efficiency and the spin efficiency was calculated to be larger than ~0.035. Second harmonic Hall measurements were carried out to estimate the charge-to-spin conversion ratio. We found that the system exhibits a large field-like-torque (spin torque efficiency ~ 0.1) and small damping-like-torque (spin torque efficiency ~0.001) compared to that reported for heavy metals. High-resolution transmission electron microscopy images show that the WTe$_{2-x}$ layer is amorphous, which may enhance the spin swapping effect by inducing large interfacial spin orbit scattering, thus contributing to a large field like torque.

Spin-orbit-torque (SOT) has been of significant research interest for high-efficiency magnetization switching, which is a prospective candidate for next generation beyond CMOS devices including ultrafast magnetoresistive random-access memories (MRAMs). [1-9] The key challenge to implement SOT-based devices is to find a material which can effectively convert between charge current and spin current. The most conventional spin current generator is the heavy metal and alloy (e.g. Ta, W, and Pt) with a relatively large spin orbit coupling (SOC). [3-5] Recently, topological materials have drawn large attention due to their potential for high spin-to-charge conversion ability. Two subclasses of topological materials that have drawn great interest are topological insulators [6,7], which have spin polarized surface states, and Weyl semimetals[8-10]. Weyl semimetals have two unique properties: (i) Weyl points induced by the strong SOC exist in the bulk not just at the interface as in a topological insulator, which means larger portion of the layer is involved in spin current generation. (ii) Specifically for type II Weyl semimetals the Fermi surface is not "point like", as in topological insulators. Crossing the bands at the fermi surface generates electron and hole pockets, [14] which leads to a much larger conductivity compared to other topological materials. Among type II Weyl semimetals, tungsten ditelluride ($WTe_2$) is a prospective candidate for generating high-efficiency spin-to-charge conversion. [9-13, 15,16] Previous research on $WTe_2$ has already shown a spin momentum locking behavior [10] and a high spin-to-charge conversion ratio. [9] However, all of the samples in these experiments were prepared by either mechanical exfoliation [10-12] or molecular beam epitaxy [9] to create the ultrathin $WTe_2$ layer, which are not the industrial application compatible for device fabrication. Therefore, research on sputtered films with similar composition remains unexplored and are highly desired.

In this work, we report our investigation on magnetron-sputtered $WTe_{2-x}$ thin films that exhibit a large room temperature (RT) field like torque (FLT). Spin pumping was used to characterize the spin-to-charge conversion, and the resulting efficiency is larger than 0.045 as defined by the ratio of charge current and spin current[17]. Additionally, second harmonic Hall measurements were carried out to characterize the charge-to-spin conversion of $WTe_{2-x}$ films, which is defined as the ratio between spin conductivity (the spin torque strength) and charge conductivity. The spin torque efficiency of FLT is ~0.1, the same magnitude as the measured spin-to-charge conversion efficiency and one magnitude larger than that of the previously reported heavy metals. [18] This is due to the amorphous structure of the $WTe_{2-x}$ layer confirmed by the scanning transmission electron microscopy (STEM) images may increase the spin orbit scattering at the interface, contributing to the large FLT. The damping like torque (DLT) is relatively small, indicating a contribution of the spin swapping effect. [19] The scanning transmission electron microscopy (STEM) images confirm the amorphous structure of the $WTe_{2-x}$ layer, which may increase the resistivity, thus increase spin orbit scattering at the interface, contributing to the large FLT.

The stack structure used for the spin pumping and SOT characterization was MgO(2)/$WTe_{2-x}$ (3, 5, 8)/CoFeB (6)/MgO (2)/Ta (2) (the unit of the thickness is in nanometers), where the MgO(2)/Ta(2) bilayer is used as a capping layer. The stack was grown at RT on thermally oxidized Si/$SiO_2$ (300 nm) substrates by a six-target Shamrock magnetron sputtering system under a base pressure better than $5\times10^{-8}$ Torr. The $WTe_{2-x}$ layer was deposited by DC power source with a $WTe_2$ composite target at 40 W with a 3 mTorr argon working pressure. The $WTe_{2-x}$ samples with different thickness are labeled as WT3, WT5, and WT8, respectively. To characterize the sample structure and elemental distribution, the cross-sectional STEM samples were prepared using an FEI Helios Nanolab G4 dual-beam focused ion beam (FIB). The STEM samples were analyzed using high angle annular dark-field (HAADF) imaging, convergent beam electron diffraction (CBED), and energy-dispersive X-ray spectroscopy. An aberration-corrected (probe-corrected)

FEI Titan G2 60-300 STEM equipped with a Super-X EDX spectrometer was used here. The STEM was operated at 200 kV with probe convergence angle of 24 mrad and a beam current of 125 pA. EDX maps were collected with 1024 pixels by 540 pixels over 43 nm by 21.5 nm areas with dwell time of 4 µs/pixel. EDX data quantification was performed at ¼ resolution using the Bruker ESPRIT 1.9 software package. The surface roughness and magnetic properties of the samples were characterized by the atomic force microscopy (AFM) and physical property measurement system (PPMS) with a vibrating sample magnetometer (VSM) model, respectively.

Fig.1 shows the structural and magnetic properties of the WT3 sample. The EDX mapping revealed the clear interfaces between each layer, as shown in Fig. 1(a). The composition of $WTe_{2-x}$ layer was characterized from the EDS measurement and the atomic ratio between W and Te elements is roughly 1:1.3. Furthermore, the STEM image (see Fig. 1(b)) suggests that there is no crystalline structure in $WTe_{2-x}$ layer. Further investigation including convergent beam electron diffraction (CBED) and selected area diffraction (SAD) did not reveal any long range order in the $WTe_{2-x}$ layer, indicating this layer is amorphous throughout, as shown in Fig. 1(c). Notably, the $WTe_{2-x}$ layer appears to be textured with local lighter and darker ~1 nm domains that may indicate local variation in W concentration.

In order to check the surface roughness of the film, we performed atomic force microscopy (AFM) to all the three samples. The AFM image of the WT3 sample is shown in Fig. 1(d). The surface root mean square (RMS) roughness is calculated to be ~0.25 nm, which is sufficient for further device fabrication and SOT study. The magnetic hysteresis (M-H) loops of the WT3 sample are plotted in Fig. 1(e), showing a magnetic anisotropy along the in-plane easy axis.

We performed spin pumping experiment to measure the spin-to-charge conversion. The samples were patterned into stripes as shown in Fig. 2(a) with a width and length of 620 and 1500 µm, respectively, using UV photo lithography and ion milling. A 55 nm thick silicon dioxide layer was deposited to insulate the wave guide from the film. The wave guides and contact pads were patterned by UV photo lithography, and a Ti (10 nm)/Au (150 nm) electrode was deposited. The wave guide of the spin pumping devices, as shown in Fig. 2(a), was similar to those in our previous reports [20-22], with the signal line width of 75 µm and ground line width of 225 µm, and separation between the lines of 37.5 µm. The illustration of the spin pumping process is shown in Fig. 2(b). The rf current generates a magnetic field, which causes the precession of the magnetization of the CoFeB layer at a gigahertz (GHz) frequency. When the frequency of magnetic field matches with the oscillation frequency of the FM layer under a certain resonance field, the spin current generated from CoFeB layer injects into the WTe$_{2-x}$ layer due to the spin pumping effect [20-24]. The injected spin current is then converted to a dc charge current due to interfacial inverse Edelstein [25,26] and bulk spin Hall effects [17,20, 27,28]. This charge current is probed by measuring the open circuit voltage $V_{total}$ of the stripe using a nanovoltmeter. The $V_{total}$ at 9 GHz for the WT3 sample is shown in Fig. 2(c). The corresponding resonance peak can be divided into a symmetric (red line) and an asymmetric (blue line) Lorentzian function part as:

$$V_{total} = \frac{V_S \Delta H^2}{\Delta H^2 + (H_{ext} - H_0)^2} + \frac{V_A (H_{ext} - H_0)}{\Delta H (\Delta H^2 + (H_{ext} - H_0)^2)} \quad (1),$$

where $\Delta H$ is the line width, $H_0$ is the resonance field, $H_{ext}$ is the applied external magnetic field, $V_S$ is the symmetric voltage component, and $V_A$ is the asymmetric voltage component. The asymmetric component is originated from anisotropic magnetoresistance (AMR) and anomalous Hall effect (AHE) of the CoFeB layer. The symmetric component is originated from spin-to-charge conversion and Seebeck effect. To remove/extract the Seebeck effect voltage ($V_{SE}$) from $V_S$ and obtain the spin-to-charge conversion voltage ($V_{SC}$), we subtract the $V_S$ at positive and negative magnetic field without changing the $V_{SE}$ by reversing the field: $V_{SC} = (V_{S(+H_0)} - V_{S(-H_0)})/2$. The resulting charge current density generated by spin to charge conversion is: $J_C = \frac{V_{SC}}{R\,w}$, where $R$ is the resistance and $w$ is the width of the stripe. To obtain the spin current density $J_S$, we measured the spin pumping signals under different frequencies $f$ and the resulting $V_{SC}$ values are shown in Fig. 2(d). The resonance frequency is fitted by the Kittel formula to get the effective saturation magnetization $M_{eff}$: $f = \frac{\gamma}{2\pi}\sqrt{H_0(H_0 + 4\pi M_{eff})}$, where $\gamma$ is the gyromagnetic ratio. To get the damping constant, the linewidth $\varDelta H$ is fitted with $f$ in the relation: $\Delta H = \Delta_0 + \frac{4\pi}{\sqrt{3}\gamma}\alpha f$, where $\varDelta_0$ is inhomogeneous line broadening factor, and $\alpha$ is the damping constant. Take WT3 sample for example, the Kittel fitting and damping constant fitting of the WT3 sample is shown in the inner part of Fig. 2(d). The resulting damping constant value is $0.0039 \pm 0.00005$ for the WT3 sample. With the damping constant, assuming that the thickness is larger than the spin diffusion length and ignoring the backflow current, we obtain the spin mixing conductance $g_{\uparrow\downarrow}$, given by:

$$g_{\uparrow\downarrow} = \frac{4\pi M_S t_{FM}}{g\mu_B}(\alpha - \alpha_0) \qquad (2),$$

where g, $\mu_B$, $t_{FM}$ and $\alpha_0$ are Landé's g-factor, Bohr magneton, ferromagnetic layer thickness and intrinsic CoFeB damping constant, respectively. The intrinsic CoFeB damping constant value we used is 0.003.[22] Thus the calculated spin mixing conductance is around $(3.3 \pm 0.2) \times 10^{18}$ m$^{-2}$ for the WT3 sample. The spin current density $J_S$ is given by:

$$J_S = \frac{g_{\uparrow\downarrow}\gamma^2 h_{rf}^2 \hbar}{8\pi\alpha^2}\frac{(4\pi M_S\gamma + \sqrt{(4\pi M_S\gamma)^2 + 4\omega^2})}{(4\pi M_S\gamma)^2 + 4\omega^2}\frac{2e}{\hbar} \qquad (3),$$

where $h_{rf}$ is the microwave rf field which can be calculated from Ampere's law, $\omega = 2\pi f$ represents the excitation frequency and $\hbar$ is the reduced Planck constant. Considering the spin diffusion in WTe$_{2-x}$ layer, the resulting spin to charge conversion efficiency $\eta$ is:

$$\eta = \frac{J_C}{J_S L \tanh(\frac{t_{WT}}{2L})} \qquad (4),$$

where $L$ is the spin diffusion length in the $WTe_{2-x}$ layer and $t_{WT}$ is the thickness of $WTe_{2-x}$ layer. Note that the $WTe_{2-x}$ layer is amorphous, which means its spin diffusion length is much smaller compared to the single crystal spin diffusion length. The spin diffusion length in single crystal $WTe_2$ is as large as 22 nm at room temperature [9]. Since $\eta$ decreases with the spin diffusion length, the single crystal value can still be used to estimate the minimum value of spin to charge conversion efficiency in our amorphous $WTe_{2-x}$ films without considering the backflow spin current. A reference sample with a stack structure MgO(2)/$WTe_{2-x}$ (5)/ MgO(2)/CoFeB (6)/MgO (2)/Ta (2) was made to measure the self-spin-pumping contribution of the CoFeB layer. The resulting voltage is ~5 µV, which will be subtracted when calculating spin to charge conversion efficiency.

At 9 GHz frequency and 2.0 V (~19.03 dBm) applied voltage, the resulting spin to charge conversion efficiencies for WT3, WT5 and WT8 samples are $\eta \geq 0.035$, 0.022 and 0.018, respectively, as shown in Fig. 2(e). The spin-to-charge conversion efficiency value is equivalent to the spin Hall angle if all of the spin-to-charge conversion generated from spin Hall effect. We found the efficiency $\eta$ is comparable with the spin Hall angle in heavy metal Ta and Pt[29], meaning even amorphous $WTe_{2-x}$ can still produce efficient spin-to-charge conversion. The decay of the spin-to-charge efficiency with the increasing of $WTe_{2-x}$ layer thickness indicates that the spin to charge conversion is dominant and likely arises from interfacial IEE rather than bulk SHE[21,22]. Increasing the $WTe_{2-x}$ layer thickness will lead to larger bulk diffusion and thus decreasing $\eta$ rather than increasing the bulk SHE and therefore increasing $\eta$. Note that the potential origin of IEE may change the pure drift diffusion model we used, thus considering this, the value of spin to charge conversion is an approximation.

To further clarify the bulk and interfacial contribution of spin charge conversion, we performed a second harmonic Hall measurement to probe the charge–to-spin conversion. The samples were patterned into Hall bars by UV photo lithography and ion milling as shown in Fig. 3(a). The length and width of the Hall bar was 100 and 10 μm, respectively. AC current with a frequency of 133 Hz and a peak value of 3 mA was applied to the channel. As we rotated the sample in xy plane from 0 to 360 degrees, the first and second harmonic Hall voltages were measured via two lock-in amplifiers. When spin current is injected into the CoFeB layer, the damping-like torque lies in-plane and has a form of: $\tau_{DL} \sim \hat{m} \times (\hat{\sigma} \times \hat{m})$ and field-like torque lies out-of-plane and has a form: $\tau_{FL} \sim \hat{\sigma} \times \hat{m}$. Assuming a 100% spin current transmission, the effective field for an in-plane damping like torque is out of the plane and has the form: $H_{DL} = \frac{\hbar \theta_{DL} J_C}{2 e M_s t_{FM}} (\hat{\sigma} \times \hat{m})$, while the effective field for the out-of-plane field-like torque is in-plane and has a form: $H_{FL} = \frac{\hbar \theta_{FL} J_C}{2 e M_s t_{FM}} [m \times (\hat{\sigma} \times \hat{m})]$, as shown in Fig. 3(b). $\theta_{DL}$ and $\theta_{FL}$ are the (effective) spin Hall angle for damping (field) like torque, respectively. As rotating the angle $\theta$, consider the initial alignment angle $\frac{\pi}{2}$, define $\varphi = \theta + \frac{\pi}{2}$, the second harmonic Hall resistance $R_{xy}^{2\omega}$ is given by[30-31,32]:

$$R_{xy}^{2\omega} = -\frac{1}{2}\left(R_{AHE}\frac{H_{DL}}{H_{ext}-H_{ani}+H_{dem}} + I\alpha\nabla T\right)\cos\varphi - R_{PHE}(2\cos^3\varphi - \cos\varphi)\frac{H_{FL}+H_{Oe}}{H_{ext}} \quad (5),$$

where the damping like torque, thermal terms such as Seebeck effect and anomalous Nernst effect gives contribution to the $\cos\varphi$ dependence, while the field-like torque and Oersted field gives contribution to the $(2\cos^3\varphi - \cos\varphi)$ dependence. The $R_{AHE}$ and $R_{PHE}$ are extracted by the first harmonic anomalous Hall effect signal and perpendicular anomalous Hall effect measurement, respectively. The total contribution of the anisotropy field and demagnetization field is around 14,000 Oe. The resistivity of layer WTe$_{2-x}$ is obtained by fitting thickness $t_{WTe}$ to $1/R_{PHE}$:

$$\frac{1}{R_{PHE}} = \frac{I_{CFB}+I_{WTe}}{V_{PHE}} = A\left(1 + \frac{\rho_{CFB}}{\rho_{WTe}t_{CFB}}t_{WTe}\right) \quad (6),$$

where $I_{CFB}$ and $I_{WTe}$ are the current flow through CoFeB and WTe$_{2-x}$ layer, respectively; $\rho$ is the resistivity and *t* is the thickness of CoFeB and WTe$_{2-x}$ layers, respectively. A is the intercept when $t_{WTe}$=0. The resulting resistivity of WTe$_{2-x}$ layer is ~350 μΩ·cm, which is about twice of the resistance of CoFeB layer. The resulting ratio can be used to calculate $J_C$ flowing in the WTe$_{2-x}$ layer. The $R_{xy}^{2\omega}$ of WT3 device is shown in Fig. 3(c), where the square red dots are the original data, the solid black line is the fitting of the original data, the blue triangles are the field-like torque and Oersted field term, and the pink triangle is the damping-like torque and thermal terms. The $H_{DL}$ and ($H_{FL}$+$H_{Oe}$) are subtracted by linear fitting the cos$\varphi$ and (2cos$^3\varphi$ − cos$\varphi$) dependent term to 1/$H_{ext}$-$H_{ani}$+$H_{dem}$ and 1/$H_{ext}$ respectively as shown in the inset of Fig. 3(c). This linear fitting will remove the thermal contribution, which remains constant under constant external magnetic field. We can see that the slope in the damping like torque term is small, which means the thermal effect is dominant in the $R_{AHE}$ term.

The resulting spin Hall angle for damping-like torque $\theta_{DL}$ and effective spin Hall angle for field-like torque $\theta_{FL}$ of all the three samples are shown in Fig. 3(d). The $\theta_{DL}$s are ~0.0007, ~0.001 and ~-0.008 for WT3, WT5 and WT8 samples. The field-like torque is much larger than the damping-like torque (~2 orders), with an effective spin Hall angle of ~-0.021, ~-0.042 and ~-0.11 respectively. The small damping like torques for all samples indicate a small spin Hall effect contribution, which is consistent with the thickness dependence of the spin pumping result. There are two unique properties of our sample set: (1) the damping like torque changes its sign when changing thickness, and (2) the field-like torque is ~100 times larger than the damping like torque. The large field-like torque is different from previous reports in heavy metals or topological materials, for which the field-like torque is either negligible[7,11,12] or smaller than 30% of damping like torque[33,34].

The small damping like torque which changes the sign may come from the competition between inhomogeneous elemental distributions. As shown in the STEM image in Fig. 1, the contrast in WTe$_{2-x}$ layer is not perfectly uniform, suggesting small domains (~1 nm) of high tungsten concentration may exist. Since tungsten is a heavy metal with a spin Hall angle ranging from 0.07 to 0.4 depending on its crystal strcutre.[29, 35,36] This may lead to a competitive scheme of spin Hall effect between the W-rich and Te-rich areas.

The large field-like torque without the existence of a damping like torque is quite rare among available literature. Since the damping like torque is small, this large field like torque cannot come from the spin Hall effect of tungsten. Shao et. al. reported a field-like torque in 2D MoS$_2$ and WSe$_2$ single crystals without the existence of a damping like torque due to enhancement of the Rashba-Edelstein effect[37]. Interfacial symmetry breaking with good crystallinity for spin generating layers is required for this enhancement. However, this explanation does not explain our observed large field like torque due to the amorphous nature of our films. Recently, the spin swapping effect[38] has been proposed[39] and potentially supported by experiments[31] to be a dominant origin of spin-orbit torques when the interfacial spin-orbit coupling is weak or when the spin diffusion length is large. As shown in Fig. 3(e), the current flows through the WTe$_{2-x}$/CoFeB bilayer. Due to the shunting effect, there is a spin polarized current flowing in the CoFeB layer, which can generate another spin current $J_{SW}$ perpendicular to the current direction due to spin swapping effect[37]. Since the WTe$_{2-x}$ layer is amorphous, the interfacial SOC is not as large as in the single crystal WTe$_2$ due to a lack of band structure as a Weyl semimetal. The small SOC has also been indicated from the small DLT in the sample. This indicates that the spin-orbit scattering of swapping current $J_{SW}$, rather than SOC, becomes the dominant scheme for generation of FLT. The perpendicular spin current $J_{SW}$ is scattered at the CoFeB/WTe$_{2-x}$ interface and is enhanced by the amorphous phase, leading to more scattering points. This scattering effect can thus produce a FLT on CoFeB layer.

In summary, we have successfully grown tungsten-tellurium compounds WTe$_{2-x}$ by magnetron sputtering. STEM imaging and CBED patterns show that the WTe$_{2-x}$ layer has an amorphous structure, and AFM image shows low roughness of the heterostructure. Spin pumping and second harmonic Hall measurements were performed to estimate the spin-to-charge and charge-to-spin conversion efficiencies. We observed a spin-to-charge conversion efficiency larger than ~0.035, mainly originating from the inverse Edelstein effect rather than the inverse spin Hall effect. The second harmonic Hall measurements also confirm that the spin Hall effect contribution is small. We observed a large field-like torque without the existence of a large damping like torque, which is unique in any reported bulk SOT materials thus far. The intrinsic nature of tungsten tellurium compounds and amorphous phase suggest that the spin swapping effect and spin-orbit scattering could be the dominant origins of the observed large field-like torque. This work broadens the ways to prepare topological materials with sufficient spin to charge conversion and produces additional evidence for the contribution of spin swapping effect to spin orbit torque.

**Acknowledgements:**

This work is supported in part by SMART, one of seven centers of nCORE, a Semiconductor Research Corporation program, sponsored by National Institute of Standards and Technology (NIST), and by UMN MRSEC program under award no. DMR-1420013. This work utilized the College of Science and Engineering (CSE) Characterization Facility at the University of Minnesota (UMN), supported in part by the NSF through the UMN MRSEC program. Portions of this work were conducted in the Minnesota Nano Center, which is supported by the National Science Foundation through the National Nano Coordinated Infrastructure Network, Award

Number NNCI-1542202. J. T. H. acknowledges support from a Doctoral Dissertation Fellowship received from the graduate school at the University of Minnesota. J. P. W. acknowledges support from Robert Hartmann Endowed Chair Professorship.

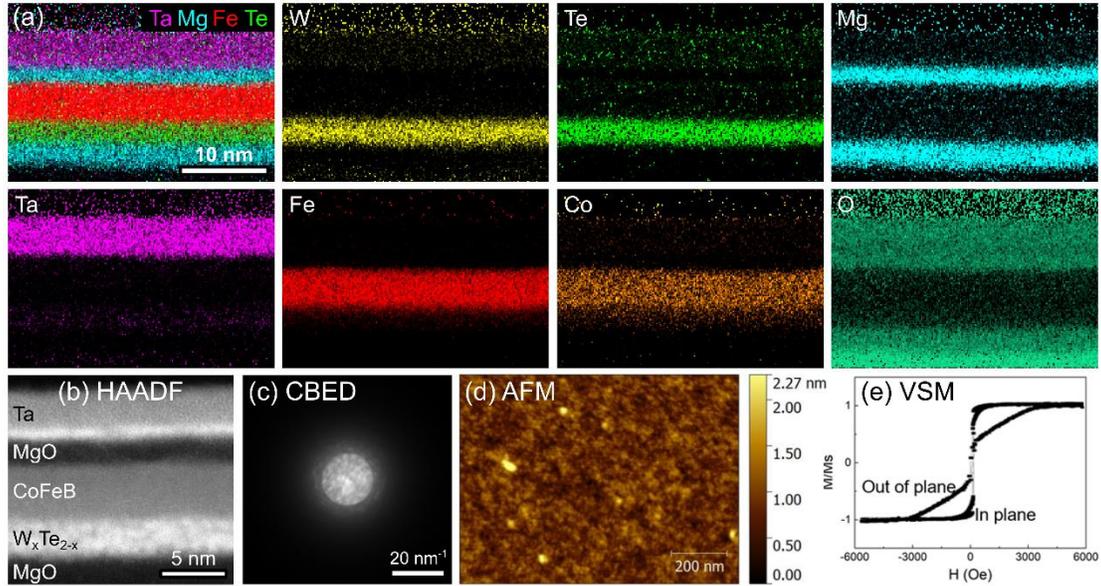

Fig. 1. (a). STEM-EDX elemental map of the MgO/WTe$_{2-x}$(3 nm)/CoFeB/MgO/Ta heterostructure grown on Si/SiO$_2$ substrate, showing the distribution of W, Te, Mg, Ta, Fe, Co, and O. A composite map of Ta, Mg, Fe and Te is also shown (top-left panel). The maps show clear definition of each layer. (b) Higher magnification HAADF-STEM image of the film structure where WTe$_{2-x}$ layer shows an amorphous structure with a nonuniform distribution of W and Te concentration. (c) A CBED pattern from the center of the WTe$_{2-x}$ layer showing no crystalline structure. (d) AFM measurement from the sample showing the film has low RMS roughness ~0.25 nm. (e) VSM measurement of the in plane and out of plane magnetization of the WT3 sample. The sample shows good in plane anisotropy.

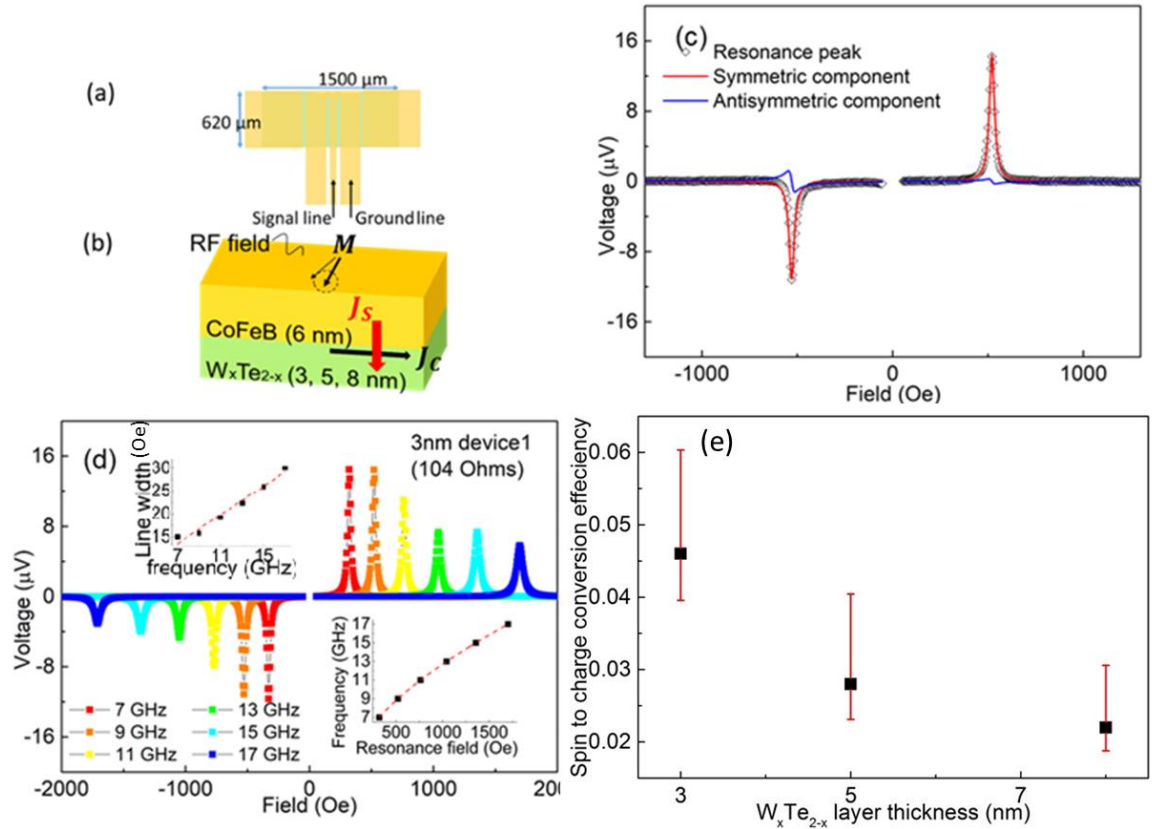

Fig. 2. (a). The schematic for spin pumping devices that are used. (b). Illustration of spin pumping. A RF current in the wave guide generates a RF field which will resonate with ferromagnetic layer and pump spin into the $WTe_{2-x}$ at the resonance field. (c). Spin pumping signal at 9GHz in the WT3 sample. The symmetric (red line) and antisymmetric (blue line) peak is subtracted. (d). Frequency dependence of the symmetric peak. The Kittel fitting and line width fitting is shown in the inner section. (e). Estimated spin to charge conversion ratio of the three samples.

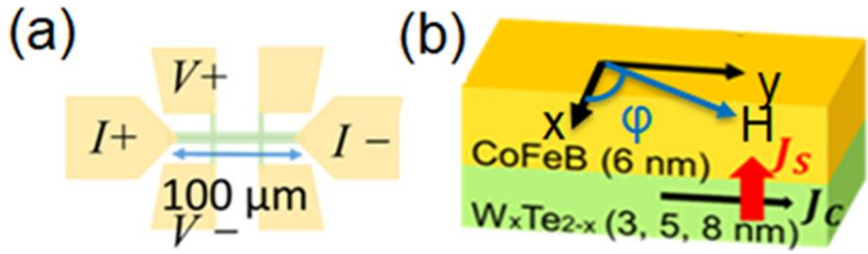

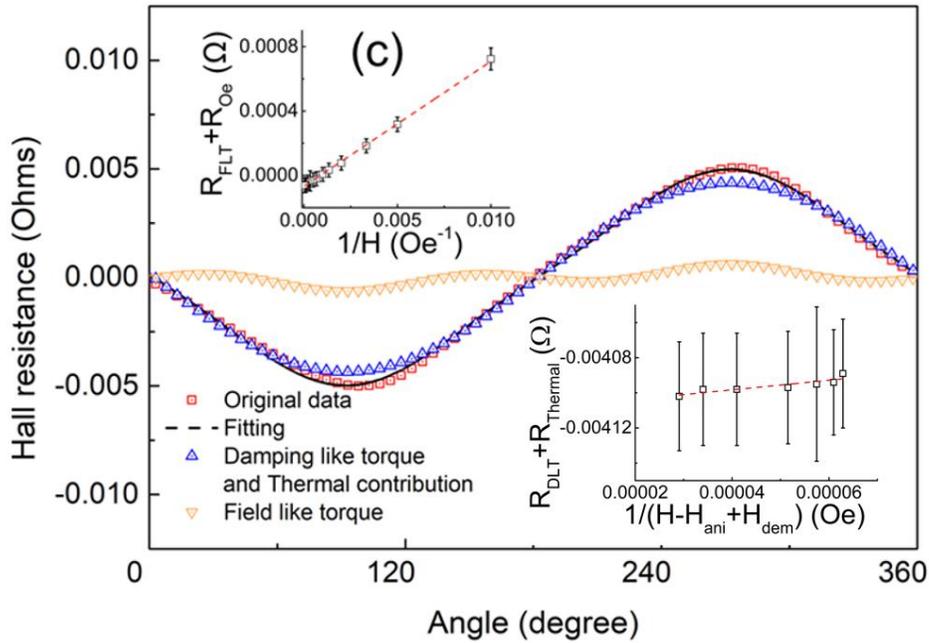

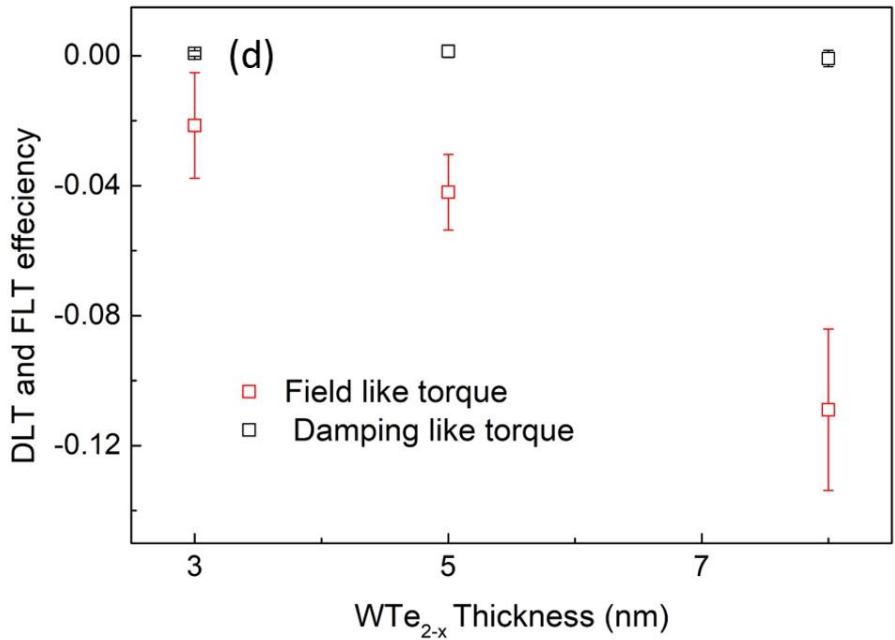

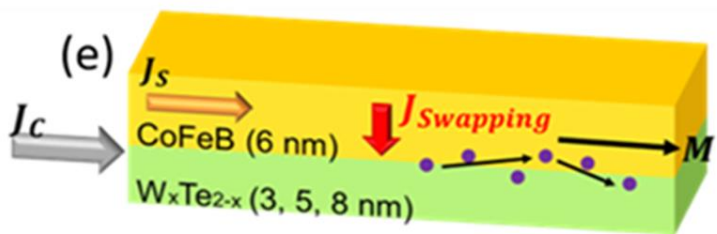

Figure 3. Figure 3. (a). The schematic for Hall bars that are used. (b). Illustration of the second harmonic Hall measurement. The spin current injected into the magnetic layer produces angular dependent first and second harmonic signals which are used for damping and field like torque analysis. (c). Second harmonic signal of WT3 sample under an external field of 100 Oe. The damping like torque (blue) and the field like torque (pink) are extracted from the original data. (d). The resulting damping like torque and the field like torque of the sample. The field like torque is two orders of magnitude larger than the damping like torque. (e). Illustration of the spin swapping effect which is a possible origin of the large damping like torque. The swapping current scattered at the interface due to spin orbit scattering which generates a torque to the ferromagnetic layer.